\documentclass[twocolumn,showpacs,preprintnumbers,amsmath,amssymb]{revtex4}

\usepackage{graphicx}
\usepackage{dcolumn}
\usepackage{bm}

\begin{document}
\preprint{Proceeding for M$^2$S-HTSC VIII meeting, July 9-14 2006,
Dresden; To appear in Physica C}

\title{Phase diagram of a dilute fermion gas with density imbalance}

\author{C.-H. Pao}
\author{S.-T. Wu}
\affiliation{%
Department of Physics, National Chung Cheng University, Chiayi 621, Taiwan}%

\author{S.-K. Yip}
\affiliation{ Institute of Physics, Academia Sinica, Nankang, Taipei 115, Taiwan}%

\date{June 15, 2006}

\begin{abstract}
We map out the phase diagram of a dilute two-component atomic
fermion gas with unequal populations and masses
 under a Feshbach resonance.
As in the case of equal masses, no uniform phase is stable for an
intermediate coupling regime. For majority component heavier, the
unstable region moves towards the BEC side.  When the coupling
strength is increased from the normal phase, there is an increased
parameter space where the transition is into the FFLO state. The
converse is true if the majority is light.
\end{abstract}


 \maketitle

Laser trapped cold dilute fermionic gas \cite{fermions} opens up a new era
to study the superfluid properties. Through the Feshbach resonance,
 the effective interaction between fermions can be
varied over a wide range such that the ground state can be tuned
from weak-coupling BCS superfluid to a strong-coupling Bose-Einstein
condensation (BEC) regime. Recent experiments \cite{ketterle} on
$^6$Li atoms with imbalance spin populations further provide another
way to probe the superfluid properties with mismatched Fermi
surfaces.

The phase diagram of this imbalanced fermion system has been studied
near the crossover region for pairing with equal masses
\cite{pao06}. Here we would like to extend our investigation to
unequal masses between the two spin species. Especially, we examine
the instabilities towards the Fulde-Ferrell-Larkin-Ovchinnikov
(FFLO) phase \cite{FFLO} and phase separated phase.
 At equal masses, the instability of
FFLO state occurs earlier for low density differences but the phase
separation reaches first for large density differences. For
increasing majority mass, the region where FFLO occurs first is
found to increase.  For majority heavier with mass ratio of $6.6$
(between $^{40}$K and $^{6}$Li), practically FFLO occurs first for
all population differences.

 We first consider the uniform phase of
 two fermion species (spin $\sigma$
 = $\uparrow$ and $\downarrow$) and mass ($m_\sigma$) in the generalized
 BCS mean field approximation,
generalizing \cite{EL} to the case of unequal populations.
 The excitation spectrum for each spin is \cite{Wu03}
 \begin{equation}
 E_{\uparrow, \downarrow} ({\bf k})\, = \, \mp \left [ {k^2 \over
 4m_r} x + h \right ] \ +\ \sqrt{
\xi({\bf k})^2 \, + \, \Delta^2}\ , \label{eqe}
\end{equation}
where $x = (m_\uparrow - m_\downarrow)/(m_\uparrow + m_\downarrow)$,
$\mu \equiv (\mu_\uparrow + \mu_\downarrow)/2$, 
$h \equiv (\mu_\uparrow - \mu_\downarrow)/2$, 
$\xi({\bf k})\, =\, \hbar^2 k^2 /(4 m_r) -
\mu$ and the reduced mass $m_r$. The scattering between fermions is
short range and can be modelled as a s-wave effective interaction
characterized by the corresponding scattering length $a$. The
pairing field $\Delta$ is then determined by
\begin{eqnarray}
 -{m_r \over 2 \pi \hbar^2 a} \Delta&
=\ \Delta\int{d^3 k \over (2 \pi)^3} \biggl [
 & {1 - f(E_\uparrow) - f( E_\downarrow) \over E_\uparrow +
E_\downarrow} \ \nonumber \\
 && \ -\ {2 m_r \over \hbar^2 k^2} \biggr ]\ ,
\label{eqg} \end{eqnarray} where $f$ is the Fermi function. We then
solve the pairing field at fixed total density $n = n_\uparrow +
n_\downarrow$ and density difference $n_d = n_\uparrow -
n_\downarrow \ge 0$ \cite{pao06}. The stable homogeneous superfluid
phase must satisfy both the superfluid density and $\partial h/
\partial n_d$ are positive \cite{pao06}.

\begin{figure}
\begin{center}
  \includegraphics*[width=6.0cm]{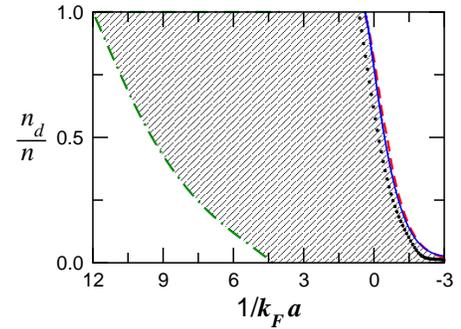}
  \end{center}
  \caption{Phase diagram for imbalanced fermion system with
  $m_\uparrow/m_\downarrow = 6.6 $.
 No uniform phase is stable
  in the shaded region (except when $n_d/n=0$ or 1).
A uniform superfluid is stable in the region left of
the dot-dashed line.
 With increasing coupling strength (moving from
  right to left), the instability lines
for the normal state (unshaded region on the
right) are:
(i) towards the FFLO (dashed lines),
(ii) phase separated state (solid line).
Dotted line represents the minimum coupling for
a solution to eq (2).
 }
  \label{fig1}
\end{figure}

Our results for $m_\uparrow/m_\downarrow = 6.6$ (the majority is
heavier) is shown in Fig \ref{fig1}. As in the equal mass case we
found that the uniform phase is unstable in the intermediate
coupling regimes (between the dotted and dot-dashed lines). Compared
with the equal mass case, this unstable region moves toward the BEC
side if the majority species is heavier.

  In the above, the dotted line is where
we first find a solution to eq (2).
However, since this $\Delta \ne 0$ solution has negative
 superfluid density
or $\partial h/ \partial n_d$
 (or both),  the system
is unstable towards
 a state with finite pairing momentum
(FFLO state)
or phase-separation.
To  find the first physical instability from the normal
state towards these phases, we
(i) solve the Cooper problem for finite wave-vector $q$, and
(ii) find the smallest coupling where the completely paired
superfluid state and the normal phase has the same
free energy.
 These results are also shown in Fig 1.
 We found that FFLO occurs earlier for
almost entire finite $n_d$ for this mass ratio.
However, a
more detailed calculation is required to determined the possible
ground state in the rest of the shaded area.

For the case with the majority is lighter, the story is quite
different. In Fig 2, we plot the phase diagram for
$m_{\uparrow}/m_{\downarrow} = 0.15$.  Compared with the equal mass
case, the unstable region moves to the BCS side.
FFLO instability occurs first only
for small $n_d/n$. For the rest of the parameter space, phase
separation occurs first.

We comment here that our results are more complete
  compared with another preprint \cite{Iskin06}, which did not
  include our FFLO and phase separation lines.

Other details will be reported elsewhere.
For a discussion of the phase diagram exactly at resonance,
see \cite{Wu06}.

In summary, we have investigated the phase diagram in an imbalanced
fermion system with unequal masses.  The phase diagram is
significantly different from its equal mass counterpart.

\begin{figure}
\begin{center}
  \includegraphics*[width=6.0cm]{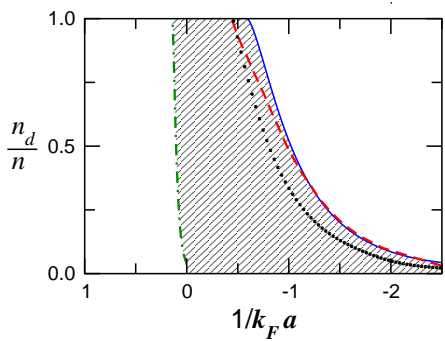}
  \end{center}
  \caption{Instability from normal state to FFLO (dashed lines) and phase
  separation (solid line) for imbalanced fermion system with
  $m_\uparrow/m_\downarrow = 0.15$. Notations are the same as Fig 1.}
  \label{fig2}
\end{figure}


\end{document}